\documentclass[twocolumn,showpacs,aps,prl,amsmath,amssymb]{revtex4}

\usepackage{epsfig}

\begin{document}

\title{Majorana zero modes in  graphene with trigonal warping} 

\author{Bal\'azs D\'ora${}^{1}$} 
\author{Mikl\'os Gul\'acsi${}^{1,2}$} 
\author{Pasquale Sodano${}^{2}$} 

\affiliation{${}^{1}$\,Max-Planck-Institut f{\"{u}}r Physik Komplexer Systeme,
N{\"{o}}thnitzer Str. 38, 01187 Dresden, Germany}
\affiliation{${}^{2}$\,Dipartimento di Fisica, Universit{\`{a}} di
Perugia, Via A. Pascoli, Perugia, 06123, Italy}

\date{23 June, 2009}

\begin{abstract}
We study the low energy properties of warped monolayer graphene, where the
symmetry of the original honeycomb lattice reveals itself. The zero energy 
solutions are Majorana fermions, whose wavefunction, originating from the 
corresponding modified Dirac equation is spatially localized. 
Experimental consequences are discussed. 
\end{abstract}
\maketitle

Graphene, a single sheet of carbon atoms have attracted enormous attention since its discovery in 2004\cite{novoselov}.
Due to the interaction of electrons with the background C-atoms, its quasiparticles obey to the Dirac equation instead of
the Schr\"odinger one, with the velocity of light ($c$) replaced by the Fermi velocity ($v_F\approx c/300$) in graphene.
This gives rise to many interesting physical phenomena, including the unconventional Hall effect,
Klein tunneling, Zitterbewegung, and peculiar electric and heat conduction\cite{review}.

In realistic graphene, there are several physical effects that appear, 
which can have a significant impact on its properties. The most important of 
these effects is the influence of ripples \cite{ripples}, leading to weak 
randomization, as scattering of short-range defects do not conserve isospin 
and valley \cite{random_ripples} and trigonal warping \cite{ando} of the 
electronic band structure which introduces asymmetry in the shape of the Fermi 
surface about each valley, as have been measured experimentally \cite{exp_warping}.

These two effects will also strongly alter the chirality of the charge carriers
in graphene \cite{ando,ryu}. Ripples, i.e., disorder leads, to a description 
of the system in terms of a Grassmannian Non Linear Sigma Model belonging to 
the symplectic symmetry class \cite{ryu}. Trigonal Warping, on the other hand, 
breaks \cite{ando} the Time reversal Symmetry  \cite{ludwig} and thus breaks 
the supersymmetry of the Non Linear Sigma Model, leading to the appearance of 
Majorana zero modes.

While much attention has been channeled into the study of ripples, there is only
a few approaches to realistic trigonal warping effects. Hence, in this letter we
want to fill this gap, by studying Majorana zero modes due to trigonal warping. 
An early computation of the effective action as sketched in \cite{ryu} shows 
that - with trigonal warping - the Non Linear Sigma Model belongs to the unitary 
symmetry class. This induces \cite{tanake} the disappearance of the perfectly 
conducting channel in a way analogous to what happens when a magnetic field is 
applied.

Hereafter, contrary to Ref. \cite{tanake}, we derive the Majorana zero modes 
from the Dirac equation, by writing the Hamiltonian, e.g., in K point, by 
explicitly adding a trigonal warping term 
$\beta (\emph{\textbf{z}}) \partial^{2}_{\emph{\textbf{z}}}$:
\begin{eqnarray}
\left(
\begin{array}{cc}
 0 & -i \partial^{}_{\emph{\textbf{z}}} + 
 \beta (\bar{\emph{\textbf{z}}}) \: \partial^{2}_{\bar{\emph{\textbf{z}}}}  \\
 i \partial^{}_{\bar{\emph{\textbf{z}}}} + 
 \beta (\emph{\textbf{z}}) \: \partial^{2}_{\emph{\textbf{z}}} & 0 \\ 
\end{array}
\right) \; . 
\label{hamiltonian}
\end{eqnarray}
Here $\emph{\textbf{z}} = \hat{k}_x + i\hat{k}_y$ and 
$\bar{\emph{\textbf{z}}} = \hat{k}_x - i \hat{k}_y$ with
the well-known notations $ \hat{k}_x = - i (\partial / \partial x)$ and 
$ \hat{k}_y = - i ( \partial / \partial y)$. 

The trigonal warping term 
$\beta (\emph{\textbf{z}}) = \beta(r) \exp( 3 i \ell \theta)$ is a 
generalized form of the second order contribution to the spectrum derived 
in Ref. \cite{ando}:
\begin{equation}
\beta \frac{a}{4 \sqrt{3}} 
e^{3 i \theta} (\hat{k}_x + i \hat{k}_x)^2 \; ,
\label{warping}
\end{equation}
where $a$ is the lattice constant, $\beta$ is an adjustable parameter, of the 
order of unity \cite{ando}, for which we introduce a slow $r$ variation, 
i.e., $\beta \equiv \beta (r)$, however we will see that the final result does 
not depend on this choice, but rather by the vorticity given by 
$\exp( 3 i \ell \theta)$.

\begin{figure}[t]
\hspace*{-2cm}
%\vspace*{-5mm}
\includegraphics[width=9.5cm]{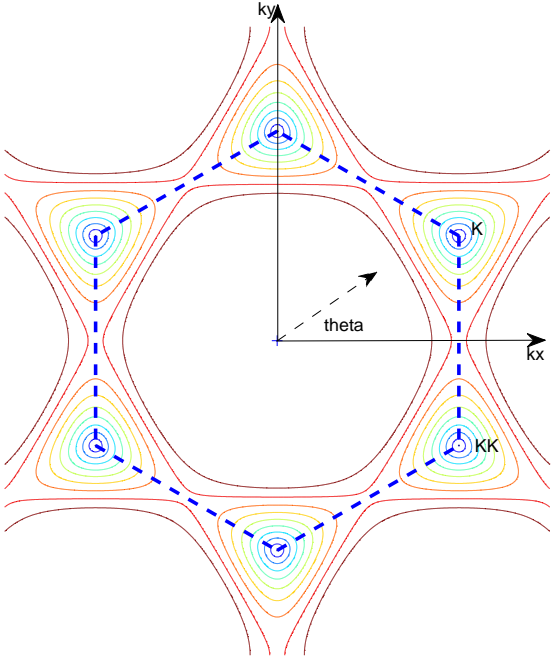} 
\caption{\label{fit1}
(Color online) The first Brillouin-zone of the honeycomb lattice (dashed lines) 
with the orientations of the coordinate axis and the equienergy surfaces close 
to the Dirac point with  trigonal warping are shown.
}
\end{figure}

The new Hamiltonian term from Eq. (\ref{warping}) is not invariant under the
time reversal operation. However, it should be noted that the total Hamiltonian
given in Eq. (\ref{hamiltonian}) does not break the original time reversal symmetry.
In fact K and K' are related to each other by time reversal. If there is not intervalley
scattering between K and K', the original time-reversal symmetry is irrelevant anyhow.

The wave-function in the K point we write as 
\begin{eqnarray}
\Psi_{K} \: = \: 
\left(
\begin{array}{c}
 u  \\
 v   \\ 
\end{array}
\right) \; ,
\end{eqnarray} 
hence the uniform limit gives back the well-known energy spectrum:
\begin{eqnarray}
&\varepsilon_{\pm} (k_x, k_y) \approx \{ k^2_x + k^2_y 
+ \beta \frac{a}{4 \sqrt{3}} \times
\nonumber \\
&[ (k^2_x - 3 k^2_y) k_x \cos 3 \theta 
+ (k^2_x - 3 k^2_y) k_y \sin 3 \theta ] \}^{1/2} \; .
\end{eqnarray}

As it can be seen, the second order term gives rise, indeed, to a trigonal warping 
of the dispersion, where the upper and lower sign corresponds to the conduction and 
valence bands, respectively. 

Going beyond the uniform limit, the full solution of the Dirac equation takes
the form: 
\begin{eqnarray}
&  - \: { \sqrt{ \beta ( r )}} \: e^{3 i \ell \phi/2} \:  
\partial^{}_{\bar{\emph{\textbf{z}}}} \: 
\left[ {\sqrt{\beta(r)}} \: e^{3 i \ell \phi/2} v
\right] \nonumber \\
& - i \partial^{}_{\emph{\textbf{z}}} u \: = \: E u \; , \nonumber \\
&  - \: { \sqrt{ \beta ( r )}} \: e^{- 3 i \ell \phi/2} \:  
\partial^{}_{\emph{\textbf{z}}} \: 
\left[ {\sqrt{\beta(r)}} \: e^{- 3 i \ell \phi/2} u
\right] \nonumber \\
& + i \partial^{}_{\emph{\textbf{z}}} v \: = \: E v \; .
\label{dirac}
\end{eqnarray}
We observe that for $\ell$ even, $\ell = 2n$ , we can completely eliminate 
the phase dependence from Eq. (\ref{dirac}) by making a transformation:
\begin{equation}
u \longrightarrow u e^{3 i n \theta} \; , 
v \longrightarrow v e^{- 3 i n \theta}
\label{transform}
\end{equation}
Hence, Eq. (\ref{dirac}) are topologically equivalent to the uniform limit, 
presented just earlier. 

The situation is completely different if $\ell$ is odd, $\ell = 2n -1$. In this
case we cannot eliminate the phase dependence. Even with the help of a
transformation (\ref{transform}) a phase factor $\exp ( 3 i \theta /2)$
always remains in Eq. (\ref{dirac}) due to which, it can be easily verified 
that,e.g., the complex conjugate of the second equation from (\ref{dirac}), 
i.e., $v*$ is identical to the first one, $u$. This guarantees one localized 
zero mode with $u = v*$. 

In the next step, we attempt to obtain the zero modes explicitly. 
For this we write the Dirac equations (\ref{dirac}) 
in compact form as: 
\begin{equation}
- i \partial^{}_{\emph{\textbf{z}}} u(r,\theta) \: + \: 
{ \sqrt{ \beta ( \emph{\textbf{z}} )     }} \: 
\partial^{}_{\bar{\emph{\textbf{z}}}} \: 
\left( {\sqrt{\beta(\emph{\textbf{z}})}} \: 
\partial^{}_{\bar{\emph{\textbf{z}}}} \right)
u(r,\theta) = 0 \; .
\end{equation}

Which, in polar coordinates becomes: 
\begin{eqnarray}
e^{i \theta} 
\left( \frac{\partial}{\partial r} - \frac{i}{r} \frac{\partial}{\partial \theta} \right)
u(r,\theta) \: + \: F(r, \theta)  
\left( \frac{\partial}{\partial r} + \frac{i}{r} \frac{\partial}{\partial \theta} \right)
\nonumber \\
\left[ F(r, \theta) 
\left( \frac{\partial}{\partial r} + \frac{i}{r} \frac{\partial}{\partial \theta} \right)
\right] u(r,\theta) \: = \: 0 \; , 
\label{ketto}
\end{eqnarray}
where, as mentioned earlier, we have odd vorticity, $\ell = 2n - 1$, 
and for simplicity we used the notation 
\begin{equation}
F(r,\theta) = f(r) e^{3 i \ell \theta / 2} e^{-i \theta} \; , 
\end{equation}
with $\beta (r) \equiv f^2 (r)$. 
 The radial solution of Eq.(\ref{ketto}) is similar to the zero modes studied 
by Ghaemi and Wilczek \cite{wilczek}: 
\begin{equation}
f^2 u^{\prime \prime} + \left( 1 + f \: f^{\prime} - \frac{n+1/2}{r} f^2 \right) 
u^{\prime} + \frac{n}{r} u = 0 \; ,
\label{harom}
\end{equation}
The solution of which for large distances, $r \gg a$, is exponentially decreasing 
\begin{equation}
u \approx \exp( - r/f^2) \; .
\end{equation}
While for $r \approx a$ tends to a constant value
\begin{equation} 
u \approx \exp \left( \frac{n a}{n + 1/2} \frac{1}{f^2} \right) \; ,
\end{equation}
hence independent of the form of $f(r)$, i.e., $\beta (r)$. 

A similar solution can be obtained for K' hence, we have two localized
Majorana zero modes in both K and K'. The number of solutions
for K equations minus the number of solutions for K' equations equal
the vorticity. This is the index theorem \cite{index1} already stated for
graphene with geometrically induced vorticity in Ref. \cite{index2}. 

This is a necessary requirement in order to have unpaired, real Majorana modes. 
In spite of the beautiful simplicity in deriving Majorana fermions, they are not 
easy to come by in nature: if the pairs of Majorana modes are not localized 
and kept apart they will immediately form a Dirac fermion. And more stringently, 
any system to exhibit Majorana modes 
has to be a perfect diamagnet, because the interaction of the complex 
fermions with photons is not diagonal. The real and imaginary components 
(which are the Majorana fermions) would rapidly mix by electromagnetic
interaction. Graphene is know to be a diamagnet \cite{diamag}, hence
Majorana fermions could exist. 

However, other systems where Majorana could be seen has to comply to these 
requirements. The easiest way to construct a perfect diamagnet is through 
a superconducting state, and hence 
other systems where Majorana zero modes have been shown to exist 
are superconducting states, such as the chiral spinless two-dimensional 
superconductors characterized by $p_x + i p_y$ order parameter \cite{supra}
and topological insulator/superconductor structures \cite{other}; or 
fractional quantum Hall states, such as the 
the $\nu = 5/2$ state \cite{moore}.  
One of the main difference between these
systems in view of the Majorana modes is that in the $\nu = 5/2$ 
quantum Hall state, a Majorana bound state is associated with charge 
e/4 quasiparticle. While in all other system, including graphene, the Majorana 
modes are electrically neutral. In the fractional quantum Hall state, thanks 
to the e/4 charge the quasiparticle's non-Abelian statistics can be probed
by measuring charge transport of the edge states \cite{exp1}. And even more
directly recent experiments \cite{exp2} have shown evidence for
the quasiparticle charge. However, detecting the Majorana fermions in 
superconductors or graphene is still lacking direct experimental
confirmation. 

Indirect confirmation of Majorana fermions in graphene may emerge from 
already well-known experimental findings. In this respect 
it is interesting to remark, firstly, that zero modes resulting from ripples, 
i.e., bond, charge, i.e., potential disorder \cite{ostrovsky1}
or random vector potential, have different properties 
from our solution, namely disorder of these types will 
suppress electron back\-scat\-te\-ring, this 
is known as the anti-localization effect \cite{altshuler}. In our
case, however, there is no suppression of backscattering, which 
leads to conventional weak localization. Hen\-ce, if weak localization
is seen to be dominating in graphene, this may well be due to topological
terms of Majorana type. 

The second possibility is related to the minimal conductivity 
of graphene \cite{conductivity}. It is believed that an intuitive explanation
of this phenomenon is that, since chirality is due to the equivalence of the
K and K' and produces non\-eq\-u\-i\-va\-lent Dirac cones at both corners of the
Brillouin zone, the wave functions of electrons have an isospin quantum
label and, thus, in isospin conserving scattering process, i.e., processes
unable to distinguish between K and K', charge carriers cannot be backscattered 
and the resistance is reduced. 

When the Non Linear Sigma Model is used to describe the transport of 
two-dimensional Dirac electrons in a random electrostatic potential, a
topological term is always required in the mathematical formulation
\cite{ryu,ostrovsky}. 
The topological term arises from Majorana fermions, and hence the
presence of Majorana zero modes can change low energy (long distance)
properties of the graphene drastically and can be responsible for its
transport properties. 

In conclusions, we have shown that taking into account trigonal warping 
explicitly,  we can account for the presence of Majorana fermions in single 
layer graphene. Experimental consequences are discussed.

\acknowledgements

The authors acknowledge fruitful discussions with Peter Fulde. PS acknowledges 
financial support of Max-Planck-Institut f\"ur Physik komplexer Systeme, Dresden.
BD was supported by the Hungarian State Grants (OTKA) K72613, and by the Bolyai
program of the HAS.

\end{document}